\documentclass[prb,twocolumn,superscriptaddress,showpacs,floatfix]{revtex4}
\usepackage{graphics}
\usepackage{graphicx}

\begin{document}
\title{Valley interference effects on a donor electron close to a Si/SiO$_2$ interface.}
\author{M.J. Calder\'on}
\affiliation{Instituto de Ciencia de Materiales de Madrid (CSIC), Cantoblanco, 28049 Madrid (Spain)}
\author{Belita Koiller}
\affiliation{Instituto de F\'{\i}sica, Universidade Federal do Rio de
Janeiro, Caixa Postal 68528, 21941-972 Rio de Janeiro, Brazil}
\author{S. {Das Sarma}}
\affiliation{Condensed Matter Theory Center, Department of Physics,
University of Maryland, College Park, MD 20742-4111}

\date{\today}

\begin{abstract}
We analyze the effects of valley interference on the quantum control and manipulation of an electron bound to a donor close to a Si/SiO$_2$ interface as a function of the valley-orbit coupling at the interface. We find that, for finite valley-orbit coupling, the tunneling times involved in shuttling the electron between the donor and the interface oscillate with the interface/donor distance in much the same way as the exchange coupling oscillates with the interdonor distance. These oscillations disappear when the ground state at the interface is degenerate (corresponding to zero valley-orbit coupling).
\end{abstract}


\maketitle

\section{Introduction}
Miniaturization of the traditional Si transistors is, after decades, still following the 
time-defying Moore's law, with a $0.7$ reduction in size every two years. Soon the effects of dopants disorder in the transistors threshold voltage will be important~\cite{shinada05} and, in a few years, their behavior will be dominated by quantum effects. These technical improvements can be used in the advancement of one of the promising candidates for the physical implementation of a quantum computer: doped Si.~\cite{kane98,vrijen00,skinner03} Si provides with a scalable framework and very long spin coherence times, very important qualities for the actual realization of a quantum computer. 

The doped Si proposal for quantum computing (QC) relies on the control and manipulation over donor electrons which shuttle between a donor site and an interface with a barrier (SiO$_2$) which is a few Bohr radii away. This process is driven by an external electric field perpendicular to the interface.  
For those relatively large distances compared to the lattice spacing, the problem may be treated within the effective mass approximation.  
Based on the single-valley effective mass approach, we have recently shown theoretically~\cite{calderonPRL06,calderon07} that a donor-bound electron can be manipulated between the donor and the interface by suitably tuning the external electric field. 
The characteristic field for which donor ionization takes place is found to be a smoothly decreasing (inversely proportional) function of the distance between the donor and the interface.~\cite{calderonPRL06} 
This result is in good agreement with earlier estimates based on tight-binding calculations where the peculiar conduction band of bulk Si with six degenerate valleys in the $\langle 100 \rangle$ directions was included,~\cite{martins04} indicating that valley interference effects may not play as strong a role in defining this characteristic field as it does, for example, in the behavior of the inter-donor exchange coupling.~\cite{koiller02PRL} 
However, other quantities of interest for the quantum control of donor electrons close to an interface may be affected by the multivalley structure of the Si conduction band, a question that deserves closer investigation.

In this paper we generalize the effective mass treatment for electrons which shuttle between a donor site and an interface by considering a two-valley structure. This is appropriate for the interface bound states, since the sixfold conduction band degeneracy in bulk Si is partially lifted at the interface, where the two valleys perpendicular to it become lower than the parallel valleys. 
Valley-orbit coupling at the interface leads to a non-degenerate ground state, where these two valleys contribute with equal weights.    
We confirm that the behavior of the characteristic field as a function of the distance of the donor to the interface is qualitatively similar to that obtained in the single-valley approximation. 
The quantitative estimates are also dependent on the magnitude of the valley-orbit coupling at the interface, a parameter which is quite sensitive to the interface type and quality.~\cite{takashina06} 
We have also studied the amplitude of the gap between the two lowest energy states at the characteristic field, a quantity related to the electron tunneling time. In this case, the single-valley estimates remain qualitatively reliable if the valley-orbit coupling at the interface may be neglected. 
However, in the presence of interface coupling, valley interference effects cause the gap to become an oscillatory function (at the lattice parameter scale) of the donor-interface distance, with quite small values of the gap attained at particular distances, leading to arbitrarily long tunneling times for donors at these positions.   
 
The outline of this paper is as follows. In Sec.~\ref{sec:model} we
introduce our model and the formalism, while results are detailed in
Sec.~\ref{sec:results}. In Sec.~\ref{sec:discussion}, we discuss the results and
conclude.

\section{Model}
\label{sec:model}
\begin{figure}
\resizebox{80mm}{!}{\includegraphics{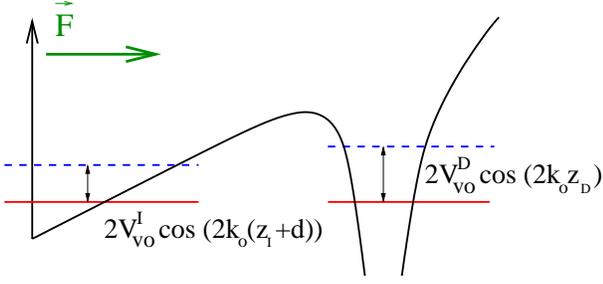}}
\caption{(Color online) Double well potential in the z-direction (Coulombic donor potential plus triangular interface/electric field potential) and the lowest levels valley-orbit splittings in the case $|V_{vo}^I| \lesssim |V_{vo}^D|$ for $F\sim F_c$. Note that the splittings depend both on the valley-orbit couplings and on the pinning phases $z_I$ and $z_D$.}
\label{fig:scheme}
\end{figure}

We consider a P donor in Si, a distance $d$ from a (001) oxide interface, under a uniform electric field perpendicular to it (see Fig.~\ref{fig:scheme}). The donor is at ${\bf r}=0$ and the interface is at ${\bf r}=(0,0,-d)$. The ground state of the electron at the donor is written~\cite{kohn55}
\begin{equation}
\Psi_D^{\rm gs}= {{1}\over{\sqrt{6}}} \sum_{\mu=1}^{6} F_D^{\mu} ({\bf r}) \phi_{\mu} ({\bf r},{\bf r}_D) \,,
\label{eq:psid}
\end{equation}
where $F_D^\mu$ are envelope functions. In the presence of an interface (which is considered to produce an infinite barrier at $z=-d$), we take~\cite{calderon07}
\begin{equation}
F_D^z = N (z+d) e^{-\sqrt{\rho^2/a^2+z^2/b^2}}
\end{equation}
with $a$ and $b$ variational parameters, and $N$ a normalization constant. Here 
$\phi_{\mu} ({\bf r},{\bf r}_D)= \Psi_{\rm Bloch}^{k_z} e^{-i {\bf k}_{\mu}\cdot {\bf r}_D} =u_{\mu} ({\bf r}) e^{i {\bf k}_{\mu}\cdot ({\bf r}-{\bf r}_D)}$ 
are the Si band edge Bloch states $(\mu = 1,2,\ldots 6)$, $|{\bf k}_{\mu}|=k_0=2\pi \,0.85/a_{Si}$ with $a_{Si}=5.4$ \AA $\,\,$  the lattice parameter. 
The reference site ${\bf r}_D$ is irrelevant for the individual Bloch states, for which it only contributes with a phase. However, for the superposition state such as given in Eq.~(\ref{eq:psid}), it represents a common pinning site of the Bloch states which, for the ground state of an isolated donor in Si, is naturally chosen to be at the position of the donor ${\bf r}_D=0$.\cite{kohn55} 

At a (001) interface, the conduction band edge six-fold degeneracy is broken~\cite{sham79,ando82,kane00} and the $z$ and $-z$ valleys are $\sim 40$ meV lower in energy than the other four valleys. 
The lowest energy states at the interface are the symmetric and antisymmetric combinations of these two valleys
\begin{equation}
\Psi_I^{\pm}= {{1}\over{\sqrt{2}}}  F_I( {\bf r}) [\phi_{z} ({\bf r},{\bf r}_I)\pm \phi_{-z} ({\bf r},{\bf r}_I)] \,,
\label{eq:interface-state}
\end{equation}
whose degeneracy is lifted by the abrupt interface valley-orbit coupling.~\cite{sham79} 
$F_I$ is the interface envelope function~\cite{calderon07}
\begin{eqnarray}
F_I &=& f(z) \times g(\rho) \nonumber\\
&=& {{\alpha^{5/2}}\over{\sqrt{12}}} (z+d)^2 e^{-\alpha (z+d)/2} 
 \times  {{\beta}\over{\sqrt{\pi}}} e^{-\beta^2 \rho^2 /2}  \,, 
\end{eqnarray} 
with $\alpha$ and $\beta$ variational parameters. In the case of an isolated well, the pinning site ${\bf r}_I$ should be near the interface, where the envelope electronic density is maximum and the potential is most attractive. Here we take ${\bf r}_I=-d$ (namely, {\em at} the interface); as shown below, the exact location of $r_I$ is not relevant for our results, as long as it is near the interface, far away from the donor site.

This problem has been previously addressed in the single valley approach.~\cite{calderonPRL06,calderon07}
In order to investigate multivalley effects in the simplest approximation, we assume that the donor ground state only involves the $z$ and $-z$ valleys: 
\begin{equation}
\Psi_D^{\pm}= {{1}\over{\sqrt{2}}}  F_D^z ({\bf r}) [\phi_{z} ({\bf r},{\bf r}_D)\pm \phi_{-z} ({\bf r},{\bf r}_D)] \,.
\label{eq:donor-state}
\end{equation}
$z$ and $-z$ are the relevant components which couple more strongly to the interface lowest states, and the lowest ones in Si under tensile strain [namely, grown on relaxed Si$_{1-x}$Ge$_x$ (001)].~\cite{koiller02PRB}
The envelopes $F_D$ and $F_I$ are the same as obtained in the single-valley calculation where the  values of the variational parameters $a$, $b$, $\alpha$, and $\beta$ were determined.
The remaining parameters, ${\bf r}_D=(0,0,z_D)$ and ${\bf r}_I=(0,0,z_I)$, are optimized variationally here.

 The Hamiltonian of a donor electron close to a Si/SiO$_2$ interface is, in rescaled atomic units  $a^*={{\hbar^2\epsilon_{Si}}/{m_\perp e^2}} =
 3.157$~nm and $Ry^*={{m_\perp e^4}/{2\hbar^2\epsilon_{Si}^2}}=
 19.98$~meV,
\begin{eqnarray}
H &=&- {{\partial^2}\over{\partial x^2}} - {{\partial^2}\over{\partial
 y^2}}- \gamma {{\partial^2}\over{\partial
 z^2}}
\nonumber \\ &+&\kappa e F z -{{2}\over{r}}+{{2Q}\over{\sqrt{\rho^2+(z+2d)^2}}}-{{Q}\over{2(z+d)}}
\nonumber \\&+& H_{vo}~,
\label{eq:h-effunits}
\end{eqnarray}
where 
$\gamma=m_\perp/ m_\|$, $Q={{(\epsilon_{\rm SiO_{2}}-\epsilon_{\rm Si})}/{(\epsilon_{\rm
 SiO_{2}}+\epsilon_{\rm Si})}}$,
$\epsilon_{\rm Si}=11.4$ and  $\epsilon_{\rm SiO_{2}}=3.8$, $\kappa=3.89 \times 10^{-7} \epsilon_{Si}^3
\left({{m}/{m_\perp}}\right)^2$ cm/kV, and the electric field $F$ is
given in kV/cm. The last term describes valley-orbit effects, namely the coupling
between different valleys due to the singular nature of both the donor
$(D)$ and the interface $(I)$ potentials. These couplings are
quantified by the parameters $V_{vo}^D$ and $V_{vo}^I$, as described
below.

In our two-valley formalism, the problem is restricted to the basis set of the lowest uncoupled donor and interface states, namely $\{F_D \Psi_{\rm Bloch}^{k_z} e^{-i k_0 z_D}$, $F_D \Psi_{\rm Bloch}^{-k_z} e^{i k_0 z_D}$, $F_I \Psi_{\rm Bloch}^{k_z} e^{-i k_0 (z_I+d)}$, $F_I \Psi_{\rm Bloch}^{k_z} e^{i k_0 (z_I+d)}\}$, leading to the Hamiltonian matrix

\begin{widetext}
\begin{equation}
{\mathbf H} =\left( \begin{array}{cccc}
 E_{D} & V_{vo}^D \cos(2 k_0 z_D) & H_{ID} e^{ i k_0 (z_D-z_I)} & 0 \\
 V_{vo}^D \cos(2 k_0 z_D) &E_{D} &  0 & H_{ID} e^{ -i k_0 (z_D-z_I)}\\
 H_{ID} e^{- i k_0 (z_D-z_I)} & 0 &E_I & V_{vo}^I \cos[2 k_0 (z_I+d)]\\
0 & H_{ID} e^{ i k_0 (z_D-z_I)}  & V_{vo}^I \cos(2 k_0 (z_I+d)) &E_I
 \end{array} \right)~,
\label{eq:2by2matrix}
\end{equation} 
\end{widetext}
where $E_{D}=\langle F_D |H| F_D \rangle$, $E_{I}=\langle F_I |H| F_I\rangle$, and $H_{ID}=\langle F_D |H| F_I \rangle$ are the same as the single valley matrix elements.  The two additional parameters, the pinning positions $z_D$ and $z_I$, are calculated variationally.
The valley-orbit (VO) coupling parameter at the donor is known from the isolated donor spectrum (splitting within the 1S manifold): 
For a P donor it is $V_{vo}^D=-1.5 $ meV.~\cite{grimmeis82,koiller02PRB} We use this value throughout. However, the VO coupling at the interface depends on the sample, and different values have been reported in the literature.~\cite{takashina06} We take $V_{vo}^I$ as a parameter varying from $0$ to $-10$ meV. The negative sign is chosen in analogy with the VO coupling at the donor, but our main conclusions only depend on the absolute value of $V_{vo}^I$. For  
$V_{vo}<0$, the ground state is a symmetric combination of the $z$ and $-z$ valleys. Full valley-orbit coupling (maximum splitting between the symmetric and antisymmetric combinations of the $z$ and $-z$ valleys, given in Eq.~(\ref{eq:interface-state}), with the symmetric one as the ground state) corresponds to $z_D=0$ at the donor well and $z_I=-d$ at the interface well.  When $V_{vo}^I=0$, the two combinations are degenerate at the interface. This degeneracy is broken when  $V_{vo}^I \cos[2 k_0 (z_I+d)] \ne 0$. Note that the donor states, defined in Eq.~\ref{eq:donor-state}, would also be degenerate if $z_D$ is such that $\cos(2 k_0 z_D)=0$.   
  
We define the overlap matrix  
\begin{widetext}
\begin{equation}
{\mathbf S}=\left( \begin{array}{cccc}
 1 & 0& S e^{ i k_0 (z_D-z_I)} & 0 \\
 0 & 1 &  0 & S e^{ -i k_0 (z_D-z_I)}\\
 S e^{- i k_0 (z_D-z_I)} & 0 &1 & 0 \\
0 & S e^{ i k_0 (z_D-z_I)}  & 0 & 1
 \end{array} \right)~,
\end{equation} 
\end{widetext}
where $S=\langle F_D | F_I \rangle$ is the overlap between the interface and donor envelopes, which is a decreasing function of $d$. The four eigenvalues $E_i$ and eigenstates $x_i$ ($i=0,1,2,3$) satisfy ${\mathbf H} x_i = E_i {\mathbf S} x_i$. 

According to our model, meaningful calculations refer to electric field intensities for which the envelope function of the donor well ground state may be described by a single 1S-like function. As shown in calculations of the Stark effect of P in Si,~\cite{debernardi06} higher donor excited states contribute to the ground state under fields above $24$ kV/cm. We thus restrict our studies to intermediate to long distances, namely $d>5.5 a^*$, so that the characteristic fields (see Fig.~\ref{fig:scheme}) remain below $\sim 24$ kV/cm.

\section{Results}
\label{sec:results}
As discussed in previous studies,~\cite{calderon07} the characteristic field $F_c$ above which the electronic ground state is at the interface, as well as the time required to `shuttle' the electron from the donor to the interface, are determined by a level anticrossing process involving the electronic ground state. The tunneling time $\tau$ has been estimated from the gap $g$ at anticrossing, $\tau =\hbar/g$. We focus here in the study of $F_c$ and $g$ as a function of $d$ and of $V_{vo}^I$, the interface VO coupling parameter.  

\begin{figure}
\resizebox{80mm}{!}{\includegraphics{main-vo1-inset.eps}}
\resizebox{80mm}{!}{\includegraphics{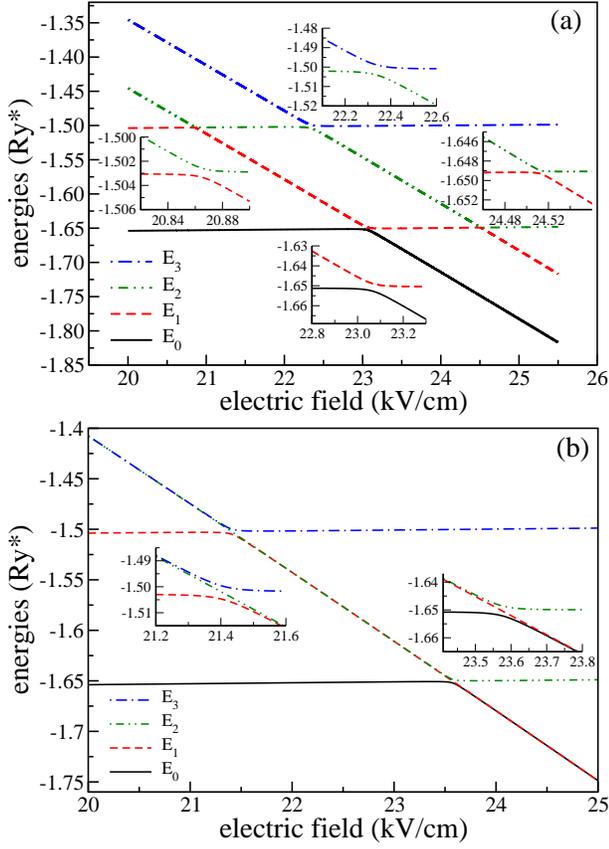}}
\caption{(Color online) (a) Here we show the eigenvalues for $d=5.94 a^*$ and $V_{vo}^I = -1 $ meV.  All the levels are coupled and there are anticrossings between the levels in pairs. The insets are blow-ups of each of the four level anticrossings. The relevant anticrossing is the one involving the ground state. 
We get qualitatively similar results for $|V_{vo}^I| \gtrsim 0.02$ meV  and for all distances $d$ not at a gap minimum (see Fig.~\ref{fig:gap-vs-d} (a)). (b) Levels dependence on the electric field when the interface VO coupling is exactly zero $V_{vo}^I=0$. We show the eigenvalues for $d=5.99 a^*$. The plot is qualitatively the same for all distances $d$. The insets are blow ups of each of the level anticrossings which shows three lines (an intermediate level remains uncoupled to other two, which display the anticrossing structure) due to the double degeneracy of the interface states. }
\label{fig:energies}
\end{figure}

In Fig.~\ref{fig:scheme} we illustrate the double well potential (donor plus interface) and the uncoupled two-valley levels (i.e., assuming $H_{ID}=0$). In the single valley approximation,~\cite{calderon07} the problem is two-dimensional, only two levels (one per well) are relevant and one anticrossing occurs at $F_c$. In the present case, for $V_{vo}^I \ne 0 $, there are two non-degenerate levels in each well and, therefore, there are four level anticrossings as illustrated for $V_{vo}^I = -1 $ meV and $d=5.94 a^*$ in Fig.~\ref{fig:energies} (a). The relevant anticrossing to determine $F_c$ and the gap $g$ is the one that involves the lowest eigenstate. When the interface states are doubly degenerate, namely $V_{vo}^I = 0$, there are only two level anticrossings involving three levels each, see Fig.~\ref{fig:energies} (b). One of the levels at the anticrossing remains uncoupled to the other two that present the typical level repulsion with an associated gap. For $0 < |V_{vo}^I| \lesssim 0.02$ meV, the results are sensitive to $d$ though the predominant behavior is similar to the degenerate case shown in Fig.~\ref{fig:energies} (b). 

\begin{figure}
\resizebox{80mm}{!}{\includegraphics{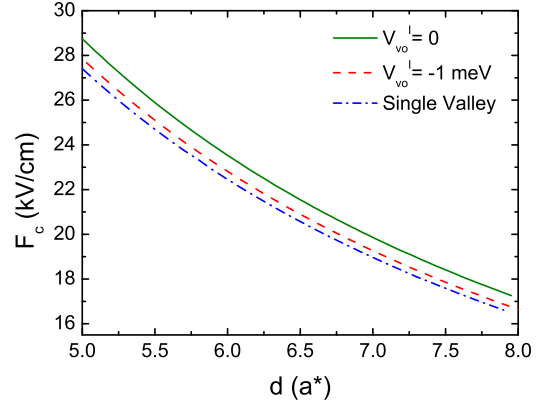}}
\caption{(Color online) $F_c$ versus $d$ for the single valley approximation, $V_{vo}^I=-1$ meV, and $V_{vo}^I=0$ (bottom to top curves). 
No oscillatory behavior is obtained as expected from the tight-binding results in Ref.~\onlinecite{martins04}. The shift between the different curves is due to the relative values of $|V_{vo}^D|$ (fixed to the bulk value $V_{vo}^D=-1.5 meV$) and $|V_{vo}^I|$.}
\label{fig:Fc-vs-d-vo1}
\end{figure}

In Fig.~\ref{fig:Fc-vs-d-vo1} we compare the values of the characteristic field $F_c$ required to move the donor electron ground state from the donor to the interface within the single valley approximation and the 2-valley formalism for $V_{vo}^I=0$ and $V_{vo}^I=-1$ meV. The behavior of $F_c$ obtained in the 2-valley formalism is qualitatively similar to the single valley approximation; only a small shift is obtained due to the different values of the ground state levels at each of the wells. The shift with respect to the single valley result is dependent on the relative values of $|V_{vo}^I|$ and $|V_{vo}^D|$, being zero if  $|V_{vo}^I|=|V_{vo}^D|$. For a particular value of $d$, $F_c$ decreases as $|V_{vo}^I|$ increases.

\begin{figure}
\resizebox{80mm}{!}{\includegraphics{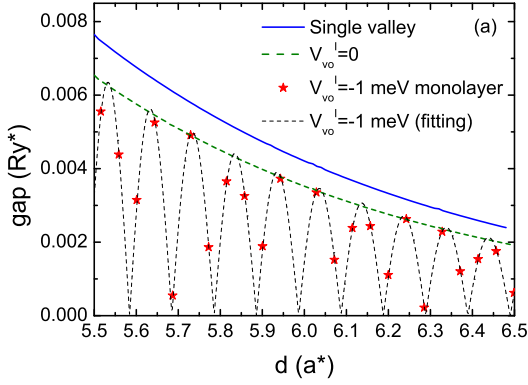}}
\resizebox{80mm}{!}{\includegraphics{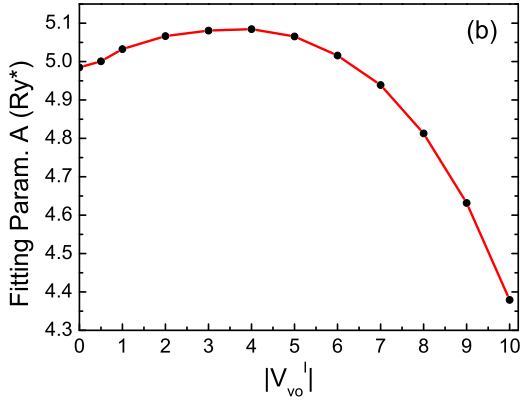}}
\caption{(Color online) (a) Gap versus $d$ for $V_{vo}^I=-1$ meV (stars and oscillating thin dashed line), $V_{vo}^I=0$ (dashed line)  and single valley approximation (solid line). 
The thin continuous dashed line is a fitting of the form ${\rm abs} [\cos(k_0 d)] \times [A \exp (-d/B)]$ with $A=5.03248 Ry^*$, and $B=0.82896 a^*$. The exponential is a fitting to the curve formed by the maxima of the oscillating function. The stars correspond to monolayer steps of $1.35$\AA. The curve for $V_{vo}^I=0$ can be also fitted to $A \exp (-d/B)$ with $A=4.98483 Ry^*$ and the same $B$. (b) Value of the fitting parameter $A$ (or 'gap' at $d=0$) as a function of $V_{vo}^I$ (in meV). The fitting parameter $B$ is always $B=0.82896 a^*$. Changing the sign of $V_{vo}^I$ leads to a shift of half a period in the gap oscillations while $A (V_{vo}^I)= A (-V_{vo}^I)$.
}
\label{fig:gap-vs-d}
\end{figure}

The smooth behavior of $F_c$ in the 2-valley formalism is not shared by the other quantity of interest: the gap at anticrossing (related to 'shuttling' tunneling times). This is illustrated in Fig.~\ref{fig:gap-vs-d} (a) where oscillations for the case $V_{vo}^I =-1$ meV are patent. The maxima of the gap oscillations correspond to tunneling times of the order of the single valley results~\cite{calderonPRL06} [see Fig.~\ref{fig:gap-vs-d} (b)]  while the gap minima would lead to much longer tunneling times. The periodicity of the oscillations is given by ${\rm abs} [\cos(k_0 d)]$ which is not commensurate with the lattice. Note that $d$ is not a continuous variable in a real system but its value changes in monolayer steps of $1.35$\AA. These values of $d$ correspond to the stars that sprinkle the continuous oscillating dashed line in Fig.~\ref{fig:gap-vs-d} (a). Therefore, tunneling times are in practice undetermined (and could be much longer than predicted in the single valley approximation) unless the position of the donor is known with atomic precision. 
The gap amplitude (defined by a phenomenological fitting discussed in the figure caption) also depends on the value of $V_{vo}^I$ as shown in Fig.~\ref{fig:gap-vs-d} (b). Increasing $|V_{vo}^I|$ leads to smaller gaps at anticrossing and therefore longer tunneling times (the difference between maximum and minimum times at a particular $d$ is $\lesssim 20\%$). In contrast, for $V_{vo}^I=0$, the gap decreases smoothly (with no oscillations) as $d$ increases.

The qualitatively distinct results obtained for $V_{vo}^I=0$ and $V_{vo}^I=-1$ meV illustrated in 
Fig.~\ref{fig:gap-vs-d} (a) emerges from the distinct level structures for these two cases, as shown in Fig.~\ref{fig:energies}.
For  $V_{vo}^I=0$, the purely interface states are degenerate (high and low field limits in this figure). Under particular fields, this degeneracy is lifted because a specific superposition couples to a donor state, leading to an anti-crossing between these levels, while a second interface state remains uncoupled and unaffected. The relevant anticrossing in our studies does not involve the uncoupled state, which is never the lowest (or highest) energy state within the anticrossing range of fields.
For $V_{vo}^I\not=0$, our numerical investigations show that for very small values of this parameter (e.g. $|V_{vo}^I| \sim 0.01$ meV) and for some values of $d$, an interface level remains uncoupled and the gap behavior is very similar to the $V_{vo}^I=0$ case. The oscillations in the gap appear consistently for all $d$ and $|V_{vo}^I| \gtrsim 0.02$ meV.

\begin{figure}
\resizebox{80mm}{!}{\includegraphics{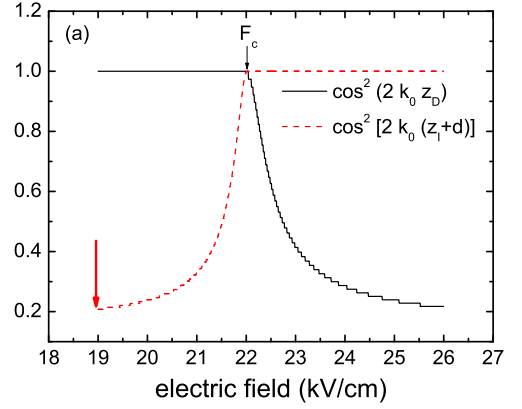}}
\resizebox{80mm}{!}{\includegraphics{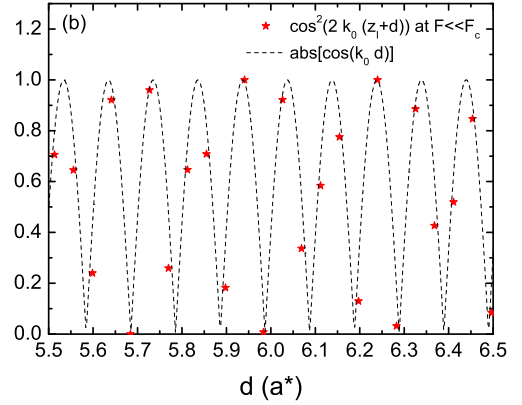}}
\caption{(Color online) Optimized variational parameters $z_I$ and $z_D$, 
related to the pinning phases, for $V_{vo}^I=-1$ meV. This figure illustrates how the oscillations in the gap emerge from
the phases in the cosines in Eq.~(\ref{eq:2by2matrix}). (a) Cosines squared of the pinning phases at $d=6.2 a^*$. The two curves cross at $F_c$. The pinning shifts from the donor at small fields (namely,  $\cos^2 (2 k_0 z_D)=1$)  to the interface  at larger fields (namely, $\cos^2 [2 k_0 (z_I+d)]=1$). The value of the cosine squared in the unoccupied well depends on $d$ and oscillates between $0$ and $1$. For instance, in the case depicted in (a) when $\cos^2 (2 k_0 z_D)=1$, $\cos^2 (2 k_0 (z_I+d))$ goes down to $\sim 0.2$ (point signaled with an arrow at the lower left corner). In (b) we plot the value of $\cos^2 (2 k_0 (z_I+d))$ at small fields ($F \ll F_c$) for different $d$'s (stars) and we see that it follows ${\rm abs} [\cos (k_0 d)] $ (continuous line). $\cos^2 (2 k_0 z_D)$ shows the same behavior when $F \gg F_c$.  These oscillations are the same shown by the gap, illustrated in Fig.~\ref{fig:gap-vs-d} (a). Therefore, for a $d$ corresponding to a gap minimum, the cosines squared go from $0$ to $1$ (and viceversa), while for a $d$ corresponding to a gap maximum, both cosines remain very close to $1$ for all electric fields. This means that for a $d$ corresponding to a gap minimum, the pinning phase shifts from donor to interface with increasing electric field while for the cases corresponding to a gap maximum each state remains pinned at its origin (donor and interface respectively) for all fields.
}
\label{fig:pinning}
\end{figure}

In Fig.~\ref{fig:pinning} we show the values of the phase pinning sites $z_I$ and $z_D$ obtained  variationally. The pinning sites depend on the applied electric field. Note that the splitting between the eigenstates in each of the wells depends on the cosines $\cos(2 k_0 z_D)$ and $\cos[2 k_0 (z_I+d)]$  (see Fig.~\ref{fig:scheme}).  Therefore, in a generic case as the one depicted in Fig.~\ref{fig:pinning} (a), the splitting between the two eigenstates (the symmetric and antisymmetric combinations of the $z$ and $-z$ valleys) depends on $F$.  Importantly, the splitting is maximum [$\cos(2 k_0 z_D)=\pm 1$ or $\cos[2k_0(z_I+d)]=\pm 1$] in the well where the electron ground state is (namely, at the donor for $F<F_c$, and at the interface for $F>F_c$). The splitting in the well where the excited state is (the interface at $F<F_c$ and the donor at $F>F_c$) oscillates with $d$ between zero and maximum with the same periodicity as the gap, as shown in Fig.~\ref{fig:pinning} (b). This implies that for the values of $d$ close to the minimum gap ($\cos(k_0 d) \sim 0$), the eigenstates at the donor (interface) are almost degenerate for $F>F_c$ ($F<F_c$). In these cases, the splitting switches from zero to its maximum possible value, and viceversa, at $F_c$, with anticrossings involving the $4$ levels and producing a minimum gap. When $d$ corresponds exactly to degenerate states [$\cos(k_0 d)=0$], the gap is larger than the ones depicted in Fig.~\ref{fig:gap-vs-d}. However, due to the incommensurability of $\cos(k_0 d)$ with respect to the lattice, the chance that $d$ satisfies $\cos(k_0 d)=0$ is very small. 

If $z_I$ and $z_D$ are fixed such that the related cosines are finite at all fields, the oscillations in the gap are preserved. The oscillations disappear when at least one of the wells has a degenerate ground state at all fields,  and the problem reduces to the single valley approach if both VO couplings are zero. 

\section{Discussions and summary}
\label{sec:discussion}
It has been known for a few years that the exchange between electrons
bound to donors separated by a
distance $R$ oscillates with $R$ due to the
interference between the valleys in the conduction band of
Si.~\cite{koiller02PRL,andres81} These oscillations present consequences for Si
quantum computing if the 2-qubit operations rely on exchange coupling.
Here we find that valley interference is also relevant for the quantum
control and manipulation of single electrons in a donor/interface
system. In particular, the tunneling times (or gaps) for the donor-interface shuttling have a periodic
(non-commensurate with the lattice) dependence on the distance $d$
between donor and interface. These oscillations arise only when
$V_{vo}^I$ is non-zero ($V_{vo}^D$ is fixed to the bulk value of
$-1.5$ meV). For $V_{vo}^I=0$ the qualitative behavior of the gap
with $d$ reduces to that of the single valley approximation, namely,
it shows no oscillations.

The oscillations in the gap versus $d$ can lead to very long tunneling
times for donors at distances such that $\cos(k_0 d) \sim 0$. In the
current experiments, donors are ion implanted and form a doped layer
with a distribution of distances with respect to the
interface.~\cite{kenton06} In this case, our results would imply that
some of those donors would never ionize in the relevant time scales,
producing an effective reduction in the planar density of donors, a
desirable feature in terms of allowing experiments to approach the
single donor behavior.~\cite{calderonPRL06}

On the other hand, arbitrarily long tunneling times for donors at
certain distances represent a serious  overhead for the final
objective of implementing a quantum computer using doped Si, as we
need to be able to manipulate the electrons many times within the spin
coherence time ($ > 1$ ms in
bulk~\cite{sousa03,witzel05,tyryshkin06}). The disappearance of the
oscillations when $V_{vo}^I \sim 0$ is of no help as a finite
valley-orbit coupling is required for a well defined qubit, in
particular, the valley-orbit splitting has to be larger than the spin
splitting, which would determine the two states of a spin qubit.
Fortunately, the statistical weight of the positions corresponding to
vanishing gaps is relatively small (see the stars in
Fig.~\ref{fig:gap-vs-d} (a)) so most donors will be at positions at
which the tunneling times are of the same order  as the ones estimated
within the single-valley approach.~\cite{calderonPRL06,calderon07}

Experimentally, different values of $V_{vo}^I$ have been reported
depending on the quality of the interface and growing
method.~\cite{takashina06} This coupling can be a complex number. We
have chosen $V_{vo}^I$ to be negative and real and explored a wide
range of values $0 \ge V_{vo}^I \ge -10$ meV consistent with the ones
reported.~\cite{takashina06} We have also checked that, for  $V_{vo}^I
> 0$, the gap versus $d$ amplitude is coincident with the case
$V_{vo}^I < 0$ though its oscillations are shifted by half a period.
We expect that a complex $V_{vo}^I$ would give different shifts in the
periodic function without affecting the gap amplitude and the decaying
behavior (given by the fitting parameters $A$ and $B$ in
Fig.~\ref{fig:gap-vs-d}).

In conclusion, we have analyzed the effect of the multivalley
structure of the Si conduction band on the manipulation of electrons
bound to donors close to a Si/SiO$_2$ interface. Our results are based on a two-valley effective mass
formalism, introduced here, and our main conclusions are expected to
remain valid if the full multivalley degeneracy at the donor (with six valleys) were included. The characteristic
field needed to shuttle an electron between the donor and the
interface well is not qualitatively affected by
valley-interference. The behavior of this
quantity is in agreement  with our previous single-valley treatment.~\cite{calderonPRL06}
However, the tunneling times involved in the
process oscillate with the distance between the donor and the interface at the
atomic scale. The single-valley results typically give a lower bound
for the tunneling times (upper bound for the anticrossing gaps). This
behavior must be taken into consideration in the practical
implementation of a doped Si quantum computer, as direct and full
control over the operations discussed here may require positioning of donors within
atomic precision.

\begin{acknowledgments}
This work is supported by LPS and NSA. B.K. also
acknowledges support from CNPq, FUJB, Millenium Institute
on Nanotechnology - MCT, and FAPERJ. M.J.C.  acknowledges support from Ram\'on y Cajal Program and MAT2006-03741 (MEC, Spain).
\end{acknowledgments}

\bibliography{multivalley}
\end{document}